%% file: abstract.tex
\documentclass[twocolumn,a4paper]{revtex4}
\usepackage[spanish,english]{babel}
\usepackage{natbib}
\usepackage{url}
\usepackage{float}

\input{macros.tex}

\begin{document}

\title{THE LARGE SCALE ORGANIZATION OF TURBULENT CHANNELS}
\author{Juan C. del \'Alamo}
\affiliation{School of Aeronautics, Plaza Cardenal Cisneros 3, 28040 Madrid,
Spain \footnote{Present address: Deptarment of Mechanical and Aerospace
Engineering, UCSD, La Jolla, CA 92093, USA. Electronic address:
\url{jalamo@ucsd.edu}}}
\author{Thesis Director: Javier Jim\'enez}

\maketitle

\section{Objectives}

We have investigated the organization and dynamics of the large turbulent
structures that develop in the logarithmic and outer layers of
high-Reynolds-number wall flows.  These structures have sizes comparable to the
flow thickness and contain most of the turbulent kinetic energy.  They produce
a substantial fraction of the skin friction and play a key role in turbulent
transport.

In spite of their significance, there is much less information about the large
structures far from the wall than about the small ones of the near-wall region.
The main reason for this is the joint requirements of large measurement records
and high Reynolds numbers for their experimental analysis.  Their theoretical
analysis has been hampered by the lack of succesful models for their
interaction with the background small-scale turbulence.

\section{Research Route and Methods}

We have performed new direct numerical simulations of turbulent channels at
higher Reynolds numbers, $180 \lesssim Re_\tau \lesssim 1900$, and in larger
computational domains than the ones previously available \citep{ala:jim:03,
ala:jim:zan:mos:04}.  They have been the first numerical experiments of wall
turbulence where the dynamics of the large scales of the logarithmic region are
properly captured. Here, $Re_\tau = u_\tau h / \nu$ is the friction Reynolds
number, based on the friction velocity $u_\tau$ and the channel half height,
$h$. The space discretization is spectral, with Fourier expansions in the
streamwise and spanwise directions ($x$ and $z$), and Chebyshev polynomials in
the wall-normal direction ($y$).  The time integration scheme is a third-order
Runge-Kutta. The resulting numerical problem, whose size is $O(10^{14})$
space-time nodes, has been solved in supercomputer centers of Spain and the
USA.  For this purpose we have developed a parallel code that has run
efficiently on up to 384 processors. 

The results from our simulations are increasingly becoming the reference
numerical data base for wall turbulence researchers.  Part of them can be
downloaded from our servers \url{http://torroja.dmt.upm.es/ftp/channels} and
\url{http://davinci.tam.uiuc.edu/data/channels}.  

The post-processing of the simulation results has adopted two complementary
approaches.  The first one has been based on statistical analysis.  We have
studied the scaling properties of the three-dimensional two-point correlation
functions of the flow variables \citep{ala:jim:03, ala:jim:zan:mos:04,
jim:ala:flo:04}.  Previous experimental and numerical studies were restricted
to only one dimension.  The second approach has been based on the analysis of
turbulent structures extracted from instantaneous flow fields.  We have
developed a novel method to isolate turbulent eddies by using a thresholding
operation with variable thresholds, which can be determined from the analogy of
a percolation transition \citep{ala:jim:zan:mos:06}.  This method has allowed
us to extract $O(10^6)$ eddies from our simulations and to analyse them in a
well-defined systematic manner, free from the subjective interpretations that
are usually inherent to structural analysis.

Finally, we have proposed a simplified model for the dynamics of the large
structures based on the linear stability equations for the turbulent mean
profile \citep{ala:jim:06, ala:flo:jim:zan:mos:06}.  The model includes an eddy
viscosity to represent the dissipation felt at the large scales because of the
smaller ones.  Because the turbulent mean profile was shown to be stable
decades ago, this problem had been abandoned by the research community.
However, we will show below that these equations have transiently-amplified
solutions that reproduce the dominant structures in channels.

\section{Outcome}

This thesis has characterized for the first time the dynamics and structure of
the self-similar range of wall turbulence.  It has given place to four papers
in the Journal of Fluid Mechanics \citep{ala:jim:zan:mos:04, jim:ala:flo:04,
ala:jim:06, ala:jim:zan:mos:06} and one in Physics of Fluids
\citep{ala:jim:03}.  Another paper is in preparation
\citep{ala:flo:jim:zan:mos:06}.

The streamwise turbulent velocity ($u$) fluctuations are organized forming very
large structures whose lengths and widths, $\lambda_x = 5-10h$ and $\lambda_z =
2-3h$, scale with the channel half height $h$ \citep{ala:jim:03}.  These
structures span the whole flow thickness in the wall-normal direction, from the
near-wall region to the center of the channel, and for this reason we have
called them {\it global modes} \citep{ala:jim:03, ala:jim:zan:mos:04}.

\begin{figure}
   \begin{center}
	    \includegraphics[width=.35\textwidth]{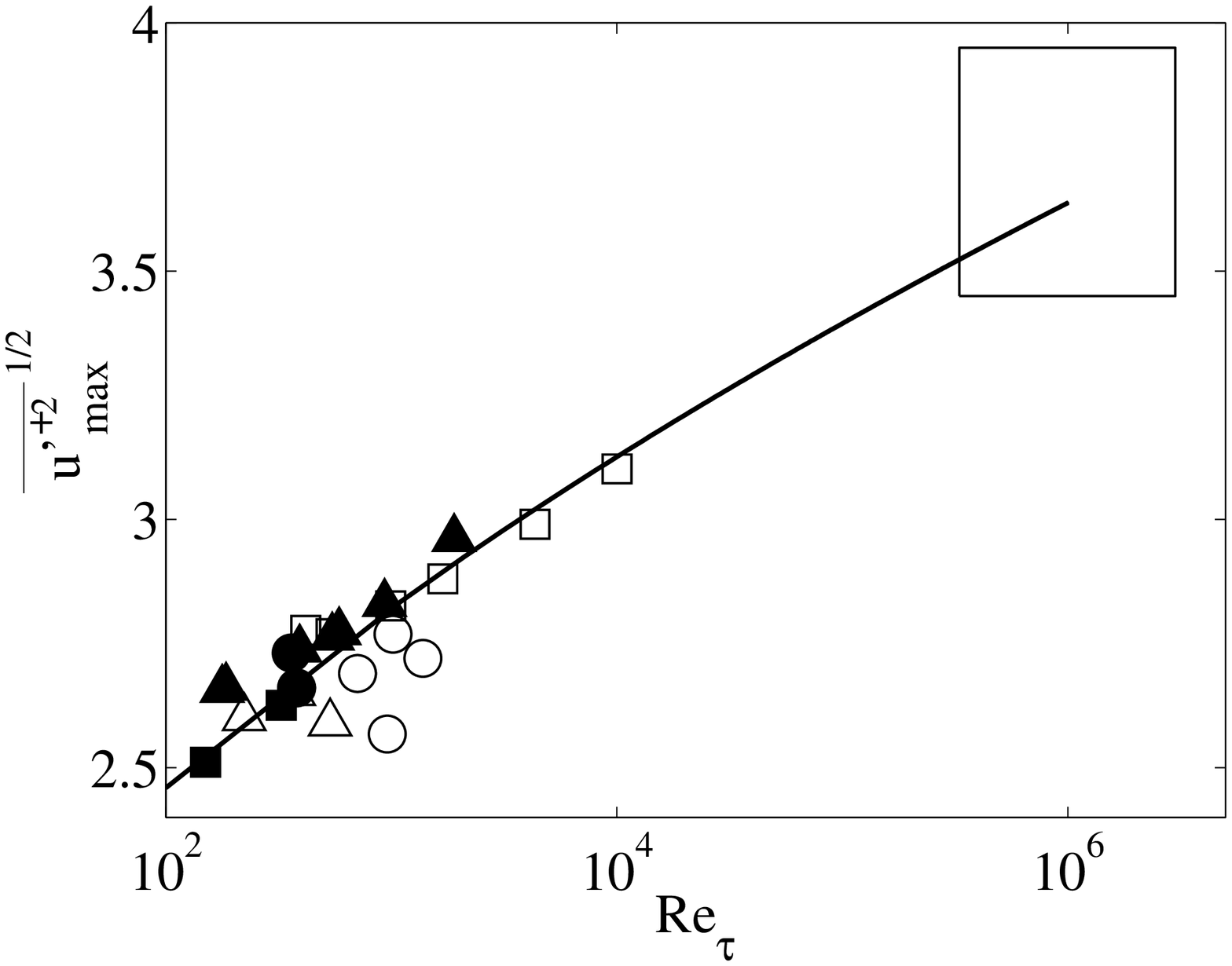}%
	    \mylab{-.27\textwidth}{.24\textwidth}{(\aaa)}%
	    \\
	    \includegraphics[width=.35\textwidth]{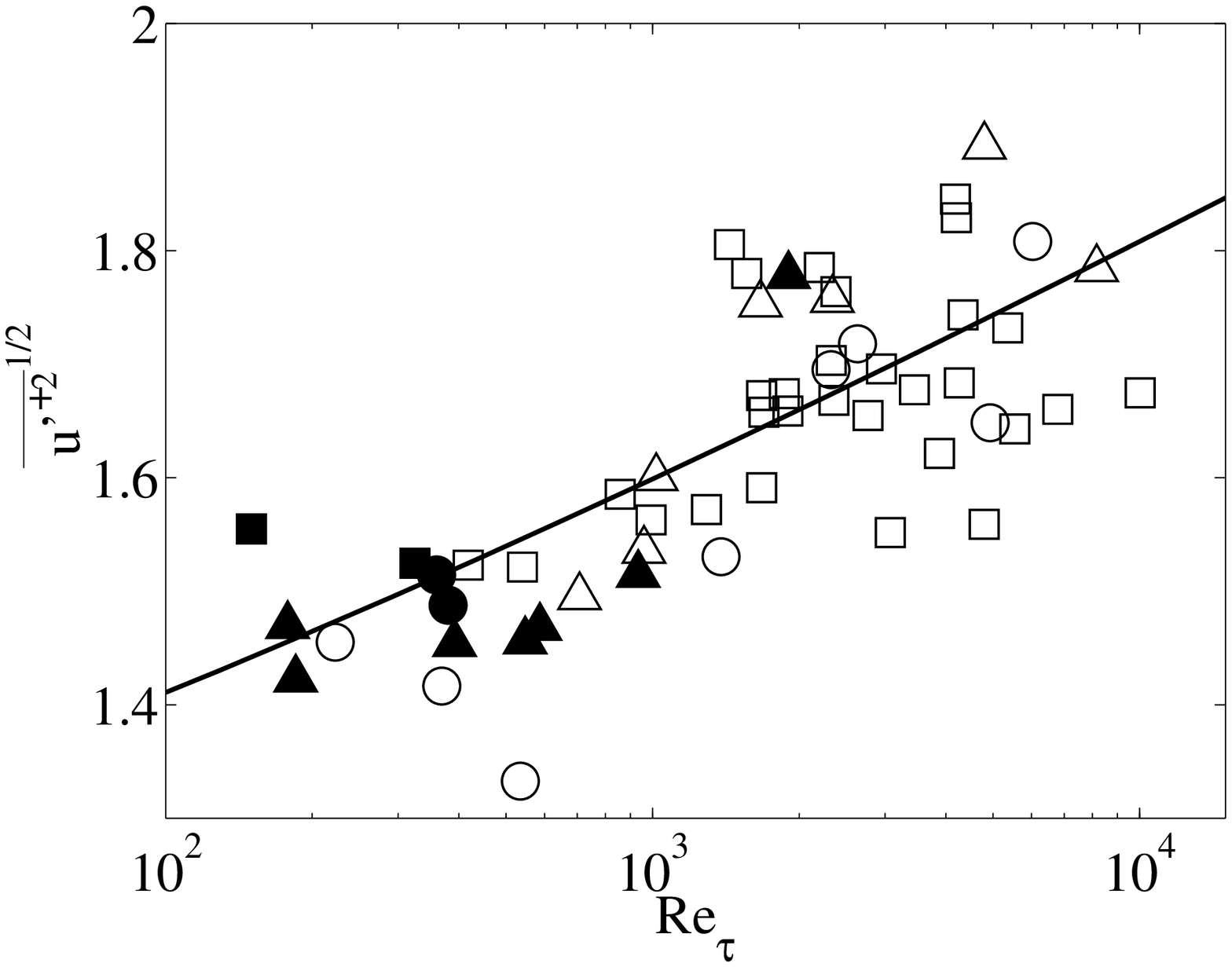}%
	    \mylab{-.27\textwidth}{.24\textwidth}{(\bbb)}%
	    \caption{(\aaa) Near-wall maximum r.m.s.
$\overline{u'^2}_{max}^{1/2+}$ of the $u$ fluctuations as a function of
$Re_\tau$.   \solid, $\overline{u'^{2}}^+_{max} \sim \log(Re_\tau)$.  (\bbb)
R.m.s.  $\overline{u'^2}^{1/2+}$ of the $u$ velocity fluctuations at $y=0.4h$
as a function of $Re_\tau$.\solid, $\overline{u'^{2}}^+ \sim \log^2(Re_\tau)$.
\squar, laboratory and atmospheric boundary layers; \trian, laboratory
channels; \circle, laboratory pipes; \solidsquar, numerical boundary layer;
\solidtrian, numerical channels; \solidcircle, numerical pipes.  The size of
the big square in (\aaa) represents the uncertainty of the atmospheric
measurements it comes from. For boundary layers $h$ is the $95\%$ thickness and
for the pipes it is the radius.}
	    \label{fig:uprimretau}
   \end{center}
\end{figure}

The influence of the global modes in the near-wall region leads to an
incomplete scaling of the turbulent energy spectra that has important
implications on the $Re_\tau$ dependence of the turbulence intensity.  While
the classical theory suggests that the turbulence intensity in the near-wall
region does not depend on $Re_\tau$ when expressed in wall units \footnote{A
variable is expressed in wall units when normalized with $u_\tau$ and $\nu$.
This is indicated with a $^+$ superscript.}, the observed incomplete scaling of
the spectra implies that it should increase as $\log(Re_\tau)$.  This
logarithmic correction explains for the first time the $Re_\tau$ behaviour of
the existing near-wall measurements of the turbulent intensity up to
atmospheric boundary layer Reynolds numbers, $Re_\tau =O(10^6)$. See figure
\ref{fig:uprimretau}(\aaa).

We have also found and explained that, far from the wall, the intensity of the
global modes scales with the mean stream velocity, $U_c$
\citep{ala:jim:zan:mos:04}.  This introduces a mixed scaling for the turbulent
kinetic energy in the outer region which tends to $U_c^2$ for $Re_\tau
\rightarrow \infty$.  This new scaling leads to a $\log^2(Re_\tau)$ correction
to the classical one in the outer region which is consistent with the available
laboratory one-point measurements, shown in figure \ref{fig:uprimretau}(\bbb).

\begin{figure}
   \begin{center}
	    \includegraphics[width=.4\textwidth]{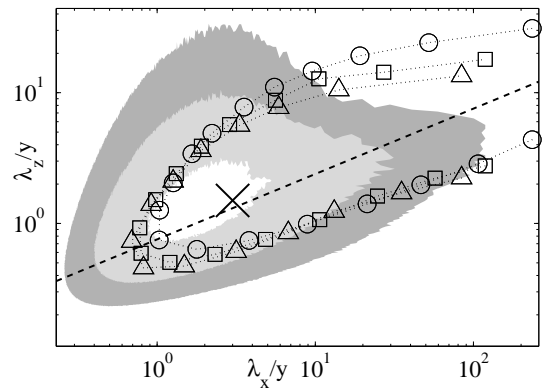}%
	    \caption{Two-dimensional spectral densities $\phi^{2D}$ as
functions of the wavelengths $\lambda_x/y$ and $\lambda_z/y$.  The data come
from our numerical turbulent channel at  $Re_\tau=950$. The line contours are
$\phi^{2D+}_u=0.1$ and each of them comes from a different wall distance
\circle, $y^+=100$; \squar, $y^+=200$; \trian, $y^+=300$.  The shaded contours
are $\phi^{2D+}_v=0.01(\times 3)0.1$, at $y^+=200$.  \dashed, $\lambda_z=
2(\lambda_x y/7)^{1/2}$.  The $\times$ marks the sizes of the vortex clusters
using the equivalence $\Delta_{x,z}=\lambda$ and $\Delta_y = y$.}
	    \label{fig:spe2D}
   \end{center}
\end{figure}
\begin{figure}
   \begin{center}
	    \includegraphics[width=.49\textwidth]{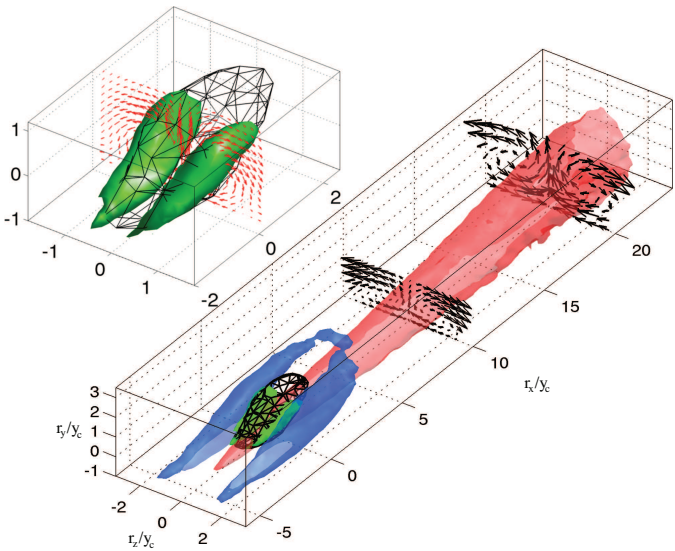}%
	    \caption{Three-dimensional plot of the average velocity field
conditioned to the vortex clusters, $\bra \vec u \ket$, in a reference frame
fixed to their centers.  The spatial coordinates, $\vec r$, are normalized with
the wall distances of the cluster centers, $y_c$.  The black mesh is an
isosurface of the p.d.f. of the vortex positions and contains $57\%$ of the
data.  The blue volume surrounding the cluster is the isosurface $\bra u'\ket^+
= 0.3$.  The red volume downstream of the cluster is the isosurface $\bra
u'\ket ^+ = -0.1$.  The green objects are vortices of the average velocity
field conditioned to the clusters. They have been obtained plotting an
isosurface of the discriminant of the velocity gradient tensor.  The arrow
plots represent $(\bra v \ket,\, \bra w \ket)$ in the planes $r_x/y_c =10,\,
20$.
	       The upper left corner shows a magnification of the surroundings
of the average position of the clusters, including a vector plot of $(\bra v
\ket,\, \bra w \ket)$ in the plane $r_x=0$.  The longest arrow measures $0.5
u_\tau$.  The data come from our numerical turbulent channel at $Re_\tau=950$.}
	    \label{fig:wakecute}
   \end{center}
\end{figure}

\begin{figure}
   \begin{center}
	    \includegraphics[width=.33\textwidth]{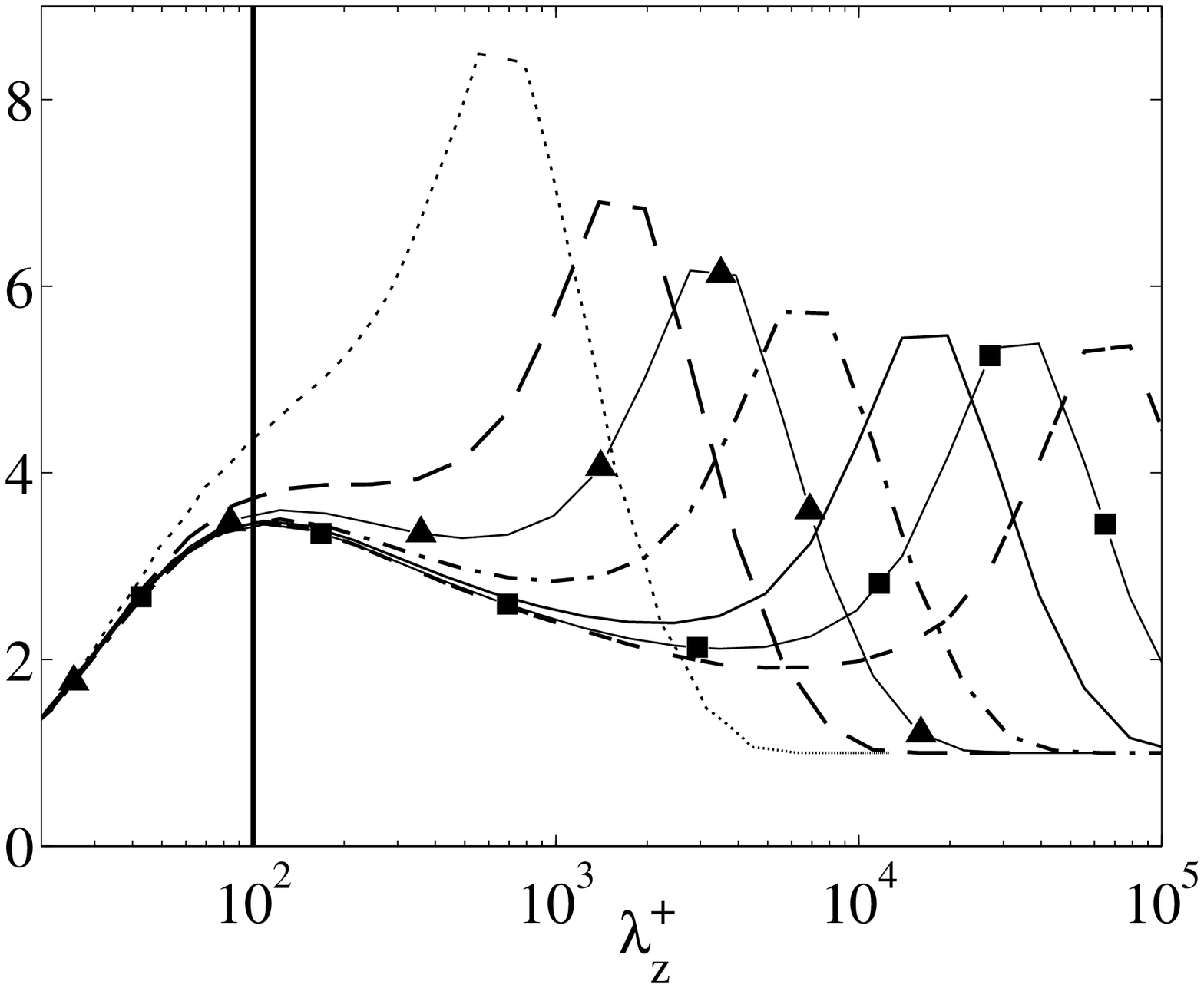}%
	    \mylab{-.3\textwidth}{.24\textwidth}{(\aaa)}%
	    \\
	    \includegraphics[width=.32\textwidth]{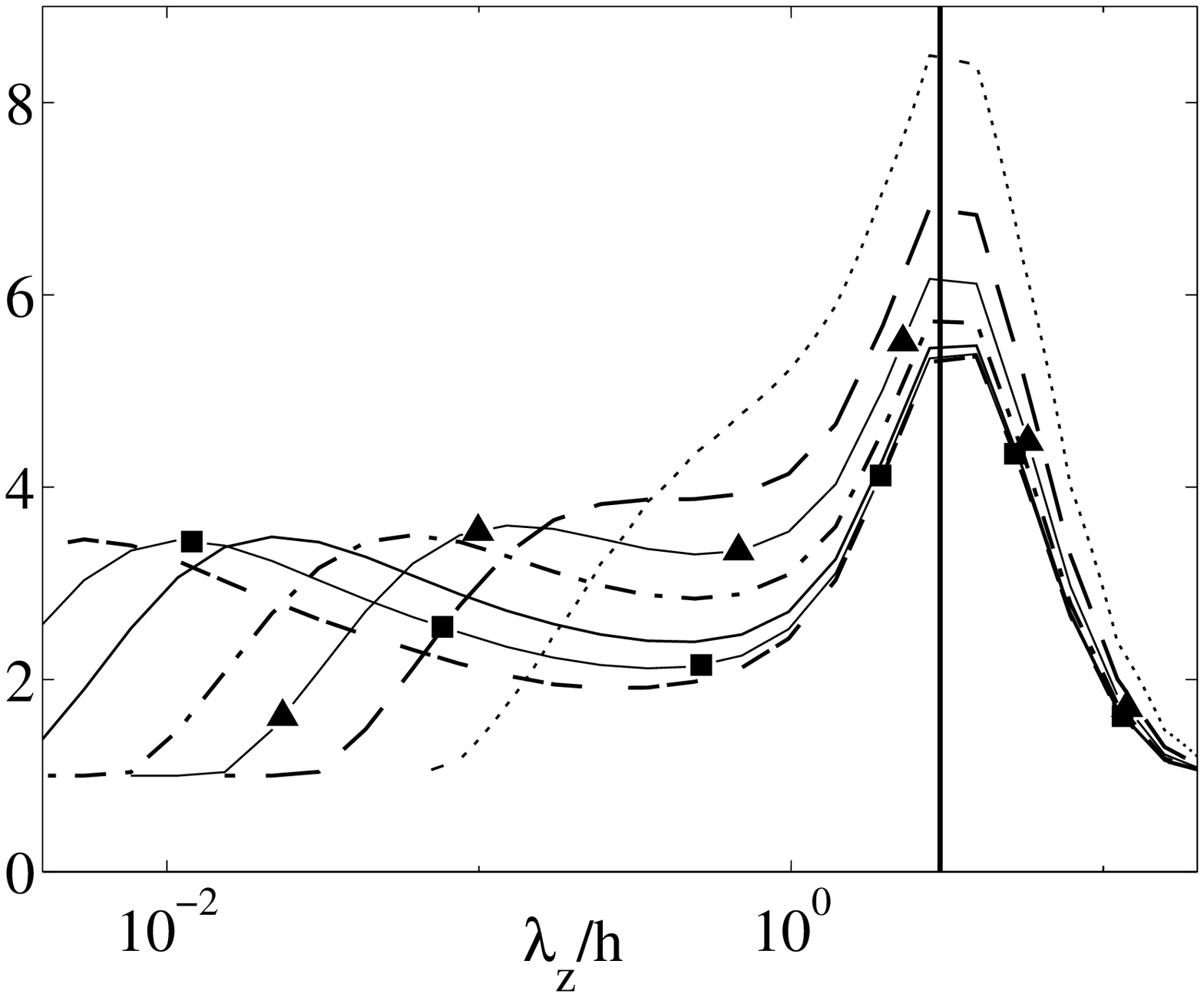}%
	    \mylab{-.29\textwidth}{.22\textwidth}{(\bbb)}%
	    \caption{Maximum energy amplifications $G(\lambda_z)$ obtained from
the linear stability equations of the mean turbulent profile for fixed
disturbance length $\lambda_x = 60h$  and different Reynolds numbers.  \dotted,
$Re_\tau=200$; \dashed, $Re_\tau=500$; \linesolidtrian, $Re_\tau=10^3$;
\chndot, $Re_\tau=2\times10^3$; \solid, $Re_\tau=5\times10^3$; \linesolidsquar,
$Re_\tau=10^4$; \dashdash, $Re_\tau = 2\times10^4$.  (\aaa) $G(\lambda_z^+)$;
the solid vertical line is $\lambda_z^+=100$.  (\bbb) $G(\lambda_z/h)$; the
solid vertical line is $\lambda_z=3h$.}
	    \label{fig:glz}
   \end{center}
\end{figure}

In the logarithmic layer, the energy spectra from our simulations reveal
important anomalies in the relation between the lengths $\lambda_x$ and widths
$\lambda_z$ of the largest turbulent structures.  Figure \ref{fig:spe2D} shows
that the form of the anomalous scaling is $\lambda_z \sim (\lambda_x y)^{1/2}$.
This suggests that the $u$ structures are ``wakes'' generated by the stirring
of the mean profile caused by compact eddies of the wall-normal ($v$) and
spanwise ($w$) velocity fluctuations \citep{ala:jim:zan:mos:04,
ala:jim:zan:mos:06}.  The value $1/2$ of the spreading exponent suggests that
the large structures are dispersed passively by the incoherent background
turbulence.  The self-similarity failure introduces further logarithmic
corrections to several ranges of the classical $k^{-1}$ spectrum of the
logarithmic region, which have been verified using data from our simulations
and from laboratory experiments at very high Reynolds numbers
\citep{ala:jim:zan:mos:04}.

The compact eddies that generate the $u$ wakes have been isolated in
instantaneous flow realizations.  They are associated to vortex clusters which
are rooted in the near-wall region and that reach very far from the wall.  On
average, these eddies consist of a wall-normal ejection surrounded by two
inclined counter-rotating vortices ,as shown in figure \ref{fig:wakecute}.
Although this average structure is consistent with a single large-scale vortex
loop, most of the individual clusters are more complex.  Their lengths and
widths ($\Delta_x$, $\Delta_z$) are proportional to their heights ($\Delta_y$)
and grow self-similarly with time after originating at different wall-normal
positions in the logarithmic region \citep{ala:jim:zan:mos:06}.  They agree
with the dominant scales of the $v$ spectrum in that region, as shown in figure
\ref{fig:spe2D}.

Figure \ref{fig:wakecute} indicates that the clusters are in fact associated to
larger structures of $u$ that appear in their wakes.  The average geometry of
these structures is a cone tangent to the wall along the $x$ axis.  The
clusters form groups of a few members within each cone, with the larger
individuals in front of the smaller ones.  This behaviour has been proven
consistent with the $\lambda_z \sim (\lambda_x y)^{1/2}$ scaling of the energy
spectrum in the logarithmic layer \citep{ala:jim:zan:mos:06}.

The clusters themselves are triggered by the wakes left by yet larger clusters
in front of them.  The whole process repeats self-similarly in a disorganized
version of the well-known vortex-streak regeneration cycle of the near-wall
region, in which the clusters and the wakes spread linearly under the effect of
the background turbulence \citep{ala:jim:zan:mos:06}.  

This linear spreading can be modelled by the Orr-Sommerfeld-Squire's equations
for the mean turbulent profile, using the eddy viscosity required to maintain
that profile \citep{ala:jim:06, ala:flo:jim:zan:mos:06}.  The dominant
structures of $u$ in turbulent channels are well described by the solutions
with the largest transient growth. Two maxima are found (see figure
\ref{fig:glz}).  The two peaks separate well when $Re_\tau$ is large enough,
and scale respectively in inner and outer units.  One corresponds to the
sublayer streaks ($\lambda_z^+=100$) and the other one to the global modes
($\lambda_z = 3h$).  The intermediate minimum is not very pronounced, and
describes self-similar modes that agree well with the observed structures of
the logarithmic layer \citep{ala:jim:06}.

The structures for the transverse velocity also agree well with the
highest-growth solutions, although they decay soon both in the linear model and
in direct simulations.  They act mainly as `seeds' for the longer-lived and
stronger structures of the streamwise velocity \citep{ala:jim:06}.

%

\end{document}

%% file: macros.tex
\usepackage[dvips]{graphicx,color}
\usepackage{amssymb,latexsym}
\usepackage{url}

\newcommand{\mylab}[3]{\raisebox{#2}[0mm][0mm]{%
\makebox[0mm][l]{\hspace*{#1}\textbf{#3}}}}
\def\spacce#1{\hskip #1pt}
\def\drawline#1#2{\raise 2.5pt\vbox{\hrule width #1pt height #2pt}}
\def\solid{\drawline{24}{.5}\nobreak}

\def\bdash{\hbox{\drawline{7}{.5}\spacce{2}}}
\def\bdashort{\hbox{\drawline{3.5}{.5}\spacce{1}}}

\def\dashed{\bdash\bdash\bdash\nobreak}
\def\dashdash{\bdashort\bdashort\spacce{3}\bdashort\bdashort\nobreak}

\def\bdot{\hbox{\drawline{1}{.5}\spacce{2}}}

\def\dotted{\hbox{\leaders\bdot\hskip 24pt}\nobreak}

\def\chndot{\hbox%
{\drawline{9.5}{.5}\spacce{2}\drawline{1}{.5}\spacce{2}\drawline{9.5}{.5}}\nobreak }
\def\circle{$\circ$\nobreak }

\def\trian{\raise 1.25pt\hbox{$\scriptstyle\triangle$}\nobreak}

\def\dtrian{\raise 1.25pt\hbox%
{$\scriptscriptstyle\bigtriangledown$}\nobreak}

\def\squar{\raise 1.25pt\hbox{$\scriptstyle\Box$}\nobreak}

\def\diamon{\raise 1.25pt\hbox{$\scriptstyle\diamond$}\nobreak}

\def\solidtrian{$\blacktriangle$\nobreak}

\def\linesolidtrian{\hbox%
{\drawline{8}{.5}\spacce{2}\solidtrian\drawline{8}{.5}}\nobreak}
\def\solidsquar{$\blacksquare$\nobreak}
\def\linesolidsquar{\hbox%
{\drawline{8}{.5}\spacce{2}\solidsquar\drawline{8}{.5}}\nobreak}
\def\solidcircle{$\bullet$\nobreak}

\def\bra{\langle}
\def\ket{\rangle}

\def\beq{\begin{equation}}
\def\eeq{\end{equation}}

\def\aaa{{\it a}}
\def\bbb{{\it b}}

\def\citalajim03{Del \'Alamo \& Jim\'enez (2003)}

